\long\def\prdfig#1#2#3
{
\begin{figure}
   \centering \leavevmode
   \epsfxsize=5.5in
   \epsfbox{#1}
   \vskip 0.3in
   \caption{\label{fig:#2} \normalsize #3}
   \hfill
\end{figure}
}
\documentstyle[aps,prl,preprint,floats,epsfig]{revtex}

\begin{document}

\preprint{\tighten\vbox{\hbox{\hfil CLNS 01/1766}
                        \hbox{\hfil CLEO 01-22}
}}

\title{\large Search for $CP$ violation in $\tau\to K \pi\nu_{\tau}$ decays.}

\author{(CLEO Collaboration)}
\date{November 29, 2001} 
\maketitle
\tighten
\begin{abstract}
We search and find no evidence for $CP$ violation in $\tau$ decays 
into the $K\pi \nu_{\tau}$ final state. 
We provide limits on the imaginary part of the coupling constant 
$\Lambda$ describing a relative contribution of the $CP$ violating processes 
with respect to the Standard Model to be $-0.172 < \Im(\Lambda) < 0.067$ at 
90\% C.L..
\end{abstract}
\newpage

{
\renewcommand{\thefootnote}{\fnsymbol{footnote}}

\begin{center}
G.~Bonvicini,$^{1}$ D.~Cinabro,$^{1}$ M.~Dubrovin,$^{1}$
S.~McGee,$^{1}$
A.~Bornheim,$^{2}$ E.~Lipeles,$^{2}$ S.~P.~Pappas,$^{2}$
A.~Shapiro,$^{2}$ W.~M.~Sun,$^{2}$ A.~J.~Weinstein,$^{2}$
% ...
\end{center}
 
\small
\begin{center}
$^{1}${Wayne State University, Detroit, Michigan 48202}\\
$^{2}${California Institute of Technology, Pasadena, California 91125}\\
% ...
\end{center}

\setcounter{footnote}{0}
}

The origin and source of $CP$ violation in fundamental fermion interactions 
is a topic of great interest. $CP$ violation has now been observed in the 
quark sector~\cite{cpk,cpk2,babar,belle}. Increasing evidence for the 
existence of neutrino masses and their mixing opens the possibility of $CP$ 
violation in the neutrino sector~\cite{marciano}. It would be odd if the mixing
effects were limited to the quarks and neutrinos only and did not appear
in the charged lepton sector. Such mixing could lead to $CP$ violation. 
There are strict limits on the mixing among the charged leptons coming from the
searches for lepton number violation~\cite{lfv}. Nevertheless,
various extensions of the Standard Model allow for the existence of $CP$
violation  not only due to the mixing but also due to the 
interference between the $W$ mediated and scalar boson mediated 
decays~\cite{mhdm,3hdm1} of the $\tau$ to the same final state. This paper 
describes a search for $CP$ violation in $\tau$ decays. 
The results of the search are interpreted within the 
context of a model where $CP$ is violated due to interference of $W$-exchange 
and an exchange of a charged scalar with complex couplings. Our previous search
for $CP$ non-conservation in correlated $\tau$ pair decay into $\pi\pi^0\nu_{\tau}$
final states~\cite{cpv_rho} benefited from the large $\tau\to \pi\pi^0\nu_{\tau}$
branching fraction yielding small statistical errors; however possible $CP$ violating 
effects are isospin-suppressed for the $\pi\pi^0\nu_{\tau}$ final 
state~\cite{tsai_kpi,kuhn}. In this search we study single
$\tau$ decays into the $K \pi\nu_{\tau}$ final state. Although this decay mode has a smaller
branching fraction, it is only supressed by the weaker $SU(3)_f$ symmetry and therefore
has greater discovery potential. A previous search using this decay was 
reported in Ref.~\cite{colin}.
\par
The most general way to search for $CP$ violation is to define a $CP$-odd
observable and then to determine its average value. A value different from 
zero would indicate $CP$ violation. Various $CP$-odd observables 
have different sensitivity to $CP$ violation. However, there is only one 
``optimal'' observable $\xi$ that has the smallest associated statistical 
error~\cite{optim1,optim2}. For a decay described by $CP$-even 
$P_{even}$ and $CP$-odd $P_{odd}$ components of the amplitude, the 
optimal variable is defined as
\begin{equation}
\label{eq:xi}
\xi = P_{odd}/P_{even}.
\end{equation}
In order to construct $\xi$ we need to know the explicit forms of $CP$-even 
and -odd parts of the amplitude in terms of experimentally measured 
parameters of the decay. This is possible only within a specific model.
Thus the choice of $\xi$ is model dependent. 
\par
We search for $CP$ violation in the decay $\tau\to K\pi\nu_{\tau}$ 
in the context of a model where the $CP$ symmetry is broken by 
an interference between Standard Model $W$-exchange and an exchange of a
scalar boson such as a charged Higgs~\cite{mhdm,3hdm1} with a complex coupling
constant $\Lambda$. We assume that $CP$ symmetry is conserved at the $\tau$ 
pair production vertex. For this model, the matrix element for the 
$\tau^-$ decay into $K\pi^-\nu_{\tau}$ final state is~\cite{tsai_kpi}
\begin{equation}
\label{eq:A}
A (\tau^{-}\to K\pi^-\nu_{\tau})\sim 
       {\bar{\it{u}}}(\nu)\gamma_{\mu}(1-\gamma_5){\it{u}}(\tau)f_V Q^{\mu} + 
\Lambda{\bar{\it{u}}}(\nu)(1+\gamma_5){\it{u}}(\tau)f_S M,
\end{equation} 
where $f_V$ and $f_S$ are the vector and the scalar form factors chosen 
to be Breit-Wigner shapes for $K^*(892)$ and $K^*_0(1430)$ resonances,
$M = 1$ GeV/$c^2$ is a constant providing a normalization of the scalar
term, and $Q^\mu$ is
\begin{equation}
\label{eq:Q}
Q^\mu = [(p_\pi - p_K)^\mu - \frac{m_\pi^2-m_K^2}
         {(p_\pi + p_K)^2}(p_\pi + p_K)^\mu].
\end{equation}
Here, $p_\pi$, $p_K$, $m_\pi$, and $m_K$ are the momenta and masses of
the outgoing pion and kaon. The square of the matrix element is
\begin{eqnarray*}
|A^{2}| & \sim & |f_V|^2(2(q \cdot Q)(Q \cdot k)-(q \cdot k)Q^2)
           + |\Lambda|^2|f_S|^2M^2(q \cdot k)
           + 2\Re(\Lambda)\Re(f_S f_V^*)M m_\tau (Q \cdot k)  
\end{eqnarray*}
\begin{equation}
\label{eq:rate}
\hspace*{-3in}       - \underline{2\Im(\Lambda)\Im(f_S f_V^*)Mm_\tau
                                                            (Q \cdot k)},
\end{equation}
where $q$ and $k$ are the 4-vectors of the $\tau$ lepton and of the neutrino,
respectively, and $m_\tau$ is the $\tau$ lepton mass.
The first three terms are $CP$ even and the last, underlined term both
violates SU(3) flavor symmetry and is $CP$ odd.
\par
To construct the optimal observable we need to express 
$(q \cdot Q)$, $(Q \cdot k)$, $Q^2$, and $(q \cdot k)$ in terms of 
experimentally measured decay parameters. From the energy and momentum 
conservation law we get
\begin{equation}
\label{eq:product1}
(q \cdot Q) = (Q \cdot k) = 
- 2([(\frac{m_\tau^2+m_H^2}{2m_H})^2 - m_\tau^2]
         [(\frac{m_H^4+(m_\pi^2-m_K^2)^2}{4m_H^2}) - m_\pi^2])^{1/2} 
         \cos\alpha, 
\end{equation}
\begin{equation}
\label{eq:product2}
Q^2 = 2m_\pi^2 + 2m_K^2 - [m_H^4+ (m_\pi^2 - m_K^2)^2]/m_H^2,
\end{equation}
\begin{equation}
\label{eq:product3}
(q\cdot k)  =  (m_\tau^2-m_H^2)/2,
\end{equation}
where $m_H$ is an invariant mass of the $(\pi K)$ system and $\alpha$
is an angle between the direction of a pion and the direction of a $\tau$ 
lepton in a pion-kaon rest frame. The angle $\alpha$ is not measurable due 
to the unknown direction of the $\tau$. However, the cosine of this angle is 
statistically equal to the product of cosines of two other measurable angles 
in the pion-kaon rest frame: $<\cos\alpha> = <\cos\beta \cos\psi>$. 
The brackets denote an averaging over the unobserved neutrino direction. 
The definitions of the angles $\beta$ and $\psi$ in terms of
measurable quantities can be found in Ref.~\cite{tsai_kpi,kuhn}.

In this study we use Eqs.~(\ref{eq:product1})-(\ref{eq:product3}) to express 
the $CP$-odd and -even parts of the squared matrix element in 
Eq.~(\ref{eq:rate}). We use the $CP$-odd and -even parts of the squared 
matrix element in Eq.~(\ref{eq:xi}) to derive the optimal observable $\xi$.
\par
The data used in this analysis were collected with the CLEO detector at the 
Cornell Electron Storage Ring (CESR) operating on or near $\Upsilon(4S)$
resonance. The data correspond to a total integrated luminosity of 
13.3 fb$^{-1}$ and contain 12.2 million $\tau^+\tau^-$ pairs. Versions 
of the CLEO detector employed here are described in Refs.~\cite{CLEOII} 
and~\cite{CLEOII5}. We estimate backgrounds by analyzing large samples 
of Monte Carlo events following the same procedures that are applied to 
the actual CLEO data. The generation of $\tau$ pair production and decay 
is modeled by the KORALB event generator~\cite{KORALB}, suitably modified
to include the charged scalar contribution to the $\tau\to K\pi\nu_{\tau}$ 
decay. The detector 
response is simulated with a GEANT-based~\cite{GEANT} Monte Carlo simulation.
\par
Tau leptons are produced in pairs in $e^+e^-$ collisions. Since the CLEO
detector is more efficient for detecting $K^0_S\to\pi^+\pi^-$ decays than for
unambiguously identified kaons, we choose to make use of the 
$\tau\to K^0_S\pi^\pm \nu_\tau$ decay, which has a 3-prong topology.
We select the candidate events on the basis of the one-vs-three topology.
The one-prong 'tag' is based on a $\tau$ candidate decaying into an electron,
muon or a single charged hadronic track, and no more than one additional 
$\pi^0$. If one prong is identified as a lepton we require the presence
of no more than one photon candidate with energy smaller than 100 MeV.
The other, 'signal' $\tau$ is required to decay into a $K^0_S$, a charged pion,
and a neutrino.
We select events with four charged tracks and zero net charge.
At CESR beam energies, the decay products of $\tau^+$ and $\tau^-$ are 
well separated in the detector. Each event is divided into two hemispheres 
by requiring one charged track to be isolated by at least $90^\circ$ from 
the other three tracks. Each track must have a momentum 
smaller than 0.85 $E_{\rm beam}$ to minimize background from Bhabha scattering 
and muon pair production. The momenta of all charged tracks are corrected 
for the energy loss in the beam pipe and tracking system. To ensure the 
existence of a $K^0_S$ decay in the three-prong hemisphere, the separation 
of the tracks 
in $z$ at the $r-\phi$ intersection must be smaller than 12 mm, the radial 
decay length of the $K^0_S$ candidate must be larger than 15 mm and the radial 
impact parameter of the $K^0_S$ must be less than 1 mm. 
\par
Background from photon conversions is suppressed by requiring the cosine of 
the angle between two tracks to be smaller than 0.99. The invariant mass of tracks
forming $K^0_S$ candidates must be within 12.5 MeV/$c^2$ of the known 
$K^0_S$ mass. To suppress background due to accidental combinations of 
the tracks we require the minimum impact parameter of the $K^0_S$ daughter tracks 
to be greater than 500 $\mu$m, $i.e.$, two times larger than the typical position 
resolution in the detector. To suppress background from $\tau\to K^0_S K\nu_{\tau}$ 
decay with the charged kaon misidentified as a pion we require $dE/dx$ information 
for the charged track accompanying the $K^0_S$ to be consistent with a pion.
\par
To estimate backgrounds coming from $\tau$ decays other than signal
we use a Monte Carlo sample containing 39.6 million $\tau^+\tau^-$ events 
in which all combinations of $\tau^+$ and $\tau^-$ decay modes are present, 
except for our signal process. Non-$\tau$ background processes include 
annihilation into multi-hadronic final states, namely 
$e^+e^-\to q\overline{q}$ ($q = u,\,d,\,s,\,c$ quarks) and 
$e^+e^-\to\Upsilon(4S)\to B\overline{B}$, as well as production of 
hadronic final states due to two-photon interactions. Backgrounds from the 
multi-hadronic physics are estimated using Monte Carlo samples which are 
slightly larger than the CLEO data and contain 42.6 million 
$q\overline{q}$ and 17.3 million $B\overline{B}$ events, respectively. 
The background due to two-photon processes is estimated 
from Monte Carlo simulation of 37 556 $2\gamma\to\tau^+\tau^-$ events, 
using the formalism of Budnev $et~al.$\cite{gamgam}. 
To study the $CP$-violating effects, we use Monte Carlo samples generated with 
and without $CP$ violation~\cite{KORALB}. The Standard Model Monte Carlo sample
contains 170 000 signal events which is four times that of the 
data, while $CP$-violating Monte Carlo samples generated with different 
values of the complex coupling $\Lambda$ consist of 200 000 events each.
\par
To suppress the background from the multi-hadronic events 
($e^+e^- \to q\bar{q}$) we require the invariant mass of the signal
hemisphere to be less than the $m_\tau$.
To suppress background from two-photon interactions we require the 
missing mass scaled with the center of mass energy to be less than 0.65 and 
the scaled transverse momentum to be greater than 0.02. We also require the 
cosine of the angle between the beam-pipe and the direction of the missing momentum
to be less than 0.95. Here, missing mass is the invariant mass of the 
difference between the 4-vector of $e^+e^-$ system and that for the 
total sum of all detected particles. 
Missing momentum is defined as a negative vector sum of all the momentum 
vectors of detected particles. The efficiency of the above selection criteria
is $(11.3\pm0.1)$\%. A total of 11 970 events have been selected from the
available CLEO data sample.
\par
After applying the above selection criteria to the Monte Carlo simulation of 
$e^+e^-\to B\bar{B}$ and $e^+e^- \to e^+e^-\gamma\gamma$ processes, 
we estimate that the remaining background from these sources contributes 
less than 0.2\% to the data sample. The background from
$e^+e^-\to q\bar{q}$ is estimated to be $(1.9\pm0.2)$\%. The dominant 
background is due to misidentified $\tau$ decays with the largest 
contribution coming from the $\tau\to K K^0 \nu_{\tau}$ decay where we
misidentify the charged kaon as a pion. We estimate this contribution to be 
$(15.2\pm1.7)$\%. The sources of background next in importance are due to 
$\tau$ decays with a $K^0_S$ in a final state with either a lost $\pi^0$,
such as $\tau\to \pi \bar{K} \pi^0 \nu_{\tau}$ [$(9.5\pm1.0)$\%] and
$\tau \to K \bar{K}^0 \pi^0 \nu_{\tau}$ [$(3.1\pm0.4)$\%], or a lost $K^0_L$ in
$\tau \to K^0 \pi \bar{K} \nu_{\tau}$ [$(8.1\pm1.4)$\%]. All other $\tau$ 
decays contribute less than 1.0\% each. The total background from 
$\tau$ decays is estimated to be $(39.2\pm2.5)$\%, and from all sources, 
$(41.3\pm2.5)\%$. As a cross check of our signal selection procedure we 
calculate a branching fraction for $\tau\to (K\pi)_{I=1/2}\nu_{\tau}$ and 
obtain a value consistent with the Particle Data Group (PDG)~\cite{pdg_kst} 
within our statistical error.
\par
$CP$ can be violated as a result of an interference between a vector 
[dominated by the $K^*(892)$] and a scalar [$e.g.$, the $K^*_0(1430)$] 
resonance in the final state.
To look for evidence of higher mass resonances we plot in 
Fig.~\ref{fig:mass} the invariant mass of the $(K^0_S\pi)$ 
system for the data, signal Monte Carlo and backgrounds. We see no evidence 
for the $K^*_0(1430)$ resonance. It can be seen in Fig.~\ref{fig:mass} that the
$K^*$ mass peak in the data is shifted by approximately $4.7\pm0.9$ MeV/$c^2$
with respect to the Monte Carlo simulation (which is based on the PDG mass
of the $K^{*+}$). This is under study, but it does not affect the results
presented in this paper.
\par
Another check for the $CP$-violating scalar 
component in the $\tau$ decay is to look at the average value of the
optimal observable as a function of the $(K^0_S\pi)$ invariant mass.
We expect the $CP$-violating effects to be maximal in the invariant
mass range laying between the resonances, $i.e.$, between 0.9 and 
1.4 GeV/$c^2$. In Fig.~\ref{fig:xi_mass} we plot $<\xi>$ separately for
$\tau^-$ and $\tau^+$ as a function of the $(K^0_S\pi)$ invariant mass for the 
data and for the Monte Carlo with maximum $CP$ violation. A difference
between the $<\xi>$ distributions for $\tau^-$ and $\tau^+$ would indicate 
$CP$ violation. We observe no difference in the $<\xi>$ distributions for the 
data and, therefore, no $CP$ violation.
\par
To calculate the limit on the $CP$ violation parameter $\Lambda$, 
we obtain the $\xi$ distribution for the data, for the Standard Model Monte 
Carlo simulation, and for the background Monte Carlo predictions. 
In Fig.~\ref{fig:data} we plot the $\xi$ distribution for both the full data
sample and for the restricted region of the $(K\pi)$ invariant mass 
0.85 GeV/$c^2 < M(K\pi) <$ 1.45 GeV/$c^2$ where the sensitivity to $CP$
violation is maximal. Here, we change the sign of $\xi$ distribution for 
the $\tau^+$ decays to add $\tau^-$ and $\tau^+$ samples together.
The corresponding average values of $<\xi>$ are listed in Table~\ref{tab:ave}.
\begin{table}[htb]
\caption {\label{tab:ave}
Average value of the optimal observable in data, Standard Model
and background Monte Carlo samples.}
\begin{center}
\begin{tabular}
{|c|c|c|} 
\multicolumn{1}{|c|} {Sample} &
\multicolumn{2}{c|} {$<\xi>$, $10^{-3}$} \\ \cline{2-3}
                   &   {Full sample}  & 
                       {0.85 GeV/$c^2 < M(K\pi) < 1.45$ GeV/$c^2$}
                                                                 \\ \hline 
data          &         $-1.5\pm 1.5$  & $-1.7\pm1.7$   \\ 
signal Monte Carlo
              &         $ 0.4\pm 1.0$  & $ 0.5\pm1.1$   \\
$\tau$ background Monte Carlo
              &         $ 0.6\pm 1.6$  & $ 0.7\pm2.3$   \\
$qq$ background Monte Carlo       
              &        $-18.1\pm14.7$  & $-23.1\pm19.1$ \\  \hline
data (background subtracted) 
              &        $-2.0\pm1.8$    & $-2.3\pm1.9$   \\ 
\end{tabular}
\end{center}
\end{table}
\par
\prdfig{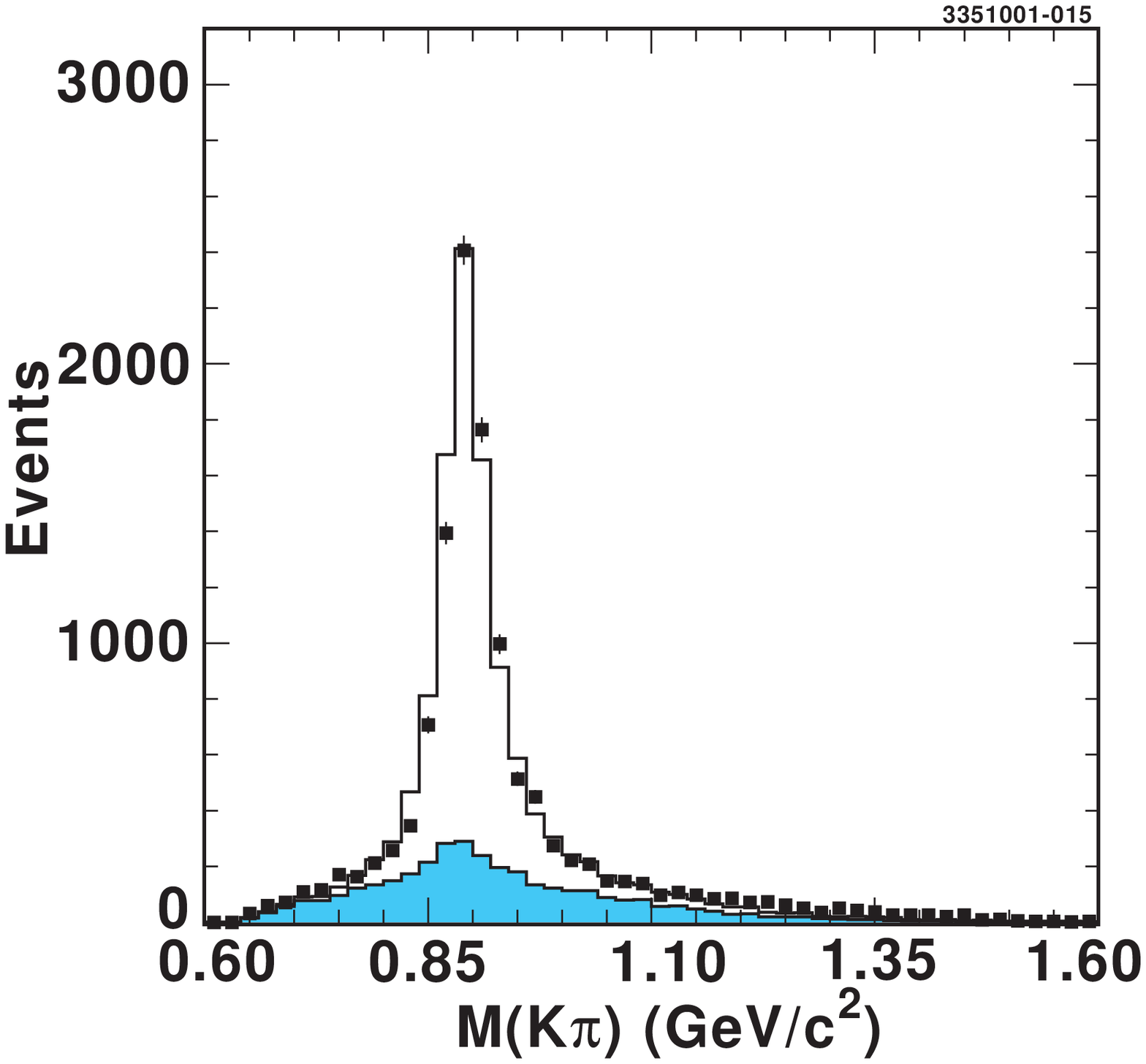}{mass}{The ($K^0_S\pi$) 
invariant mass for data (squares), signal Monte Carlo prediction (solid line) 
and background (hatched histogram).}
\prdfig{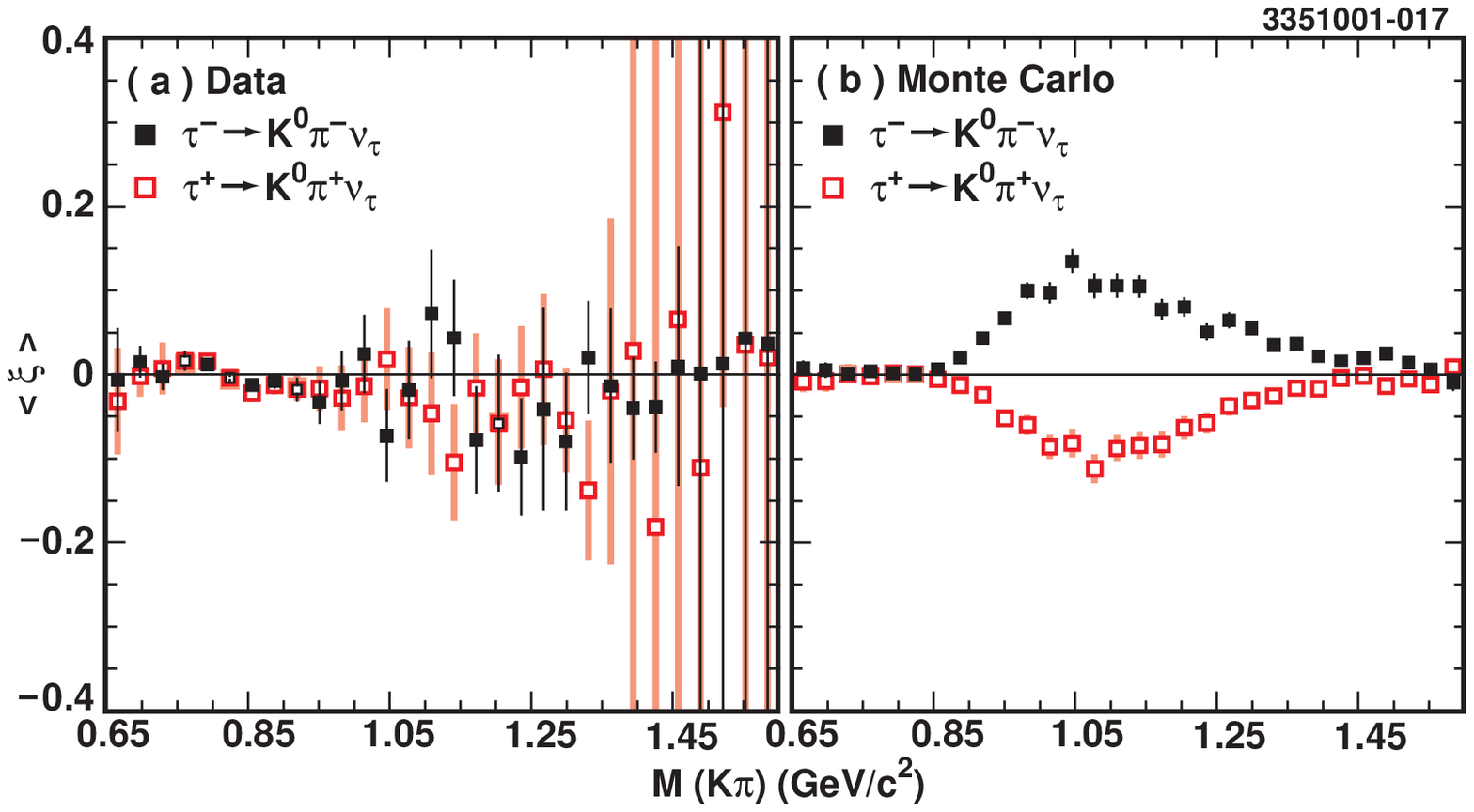}{xi_mass}{Average value of the
optimal observable as a function of the $(K^0_S\pi)$ invariant mass for
(a) data and (b) Monte Carlo with maximum $CP$ violation $\Im(\Lambda)=1$.}
\prdfig{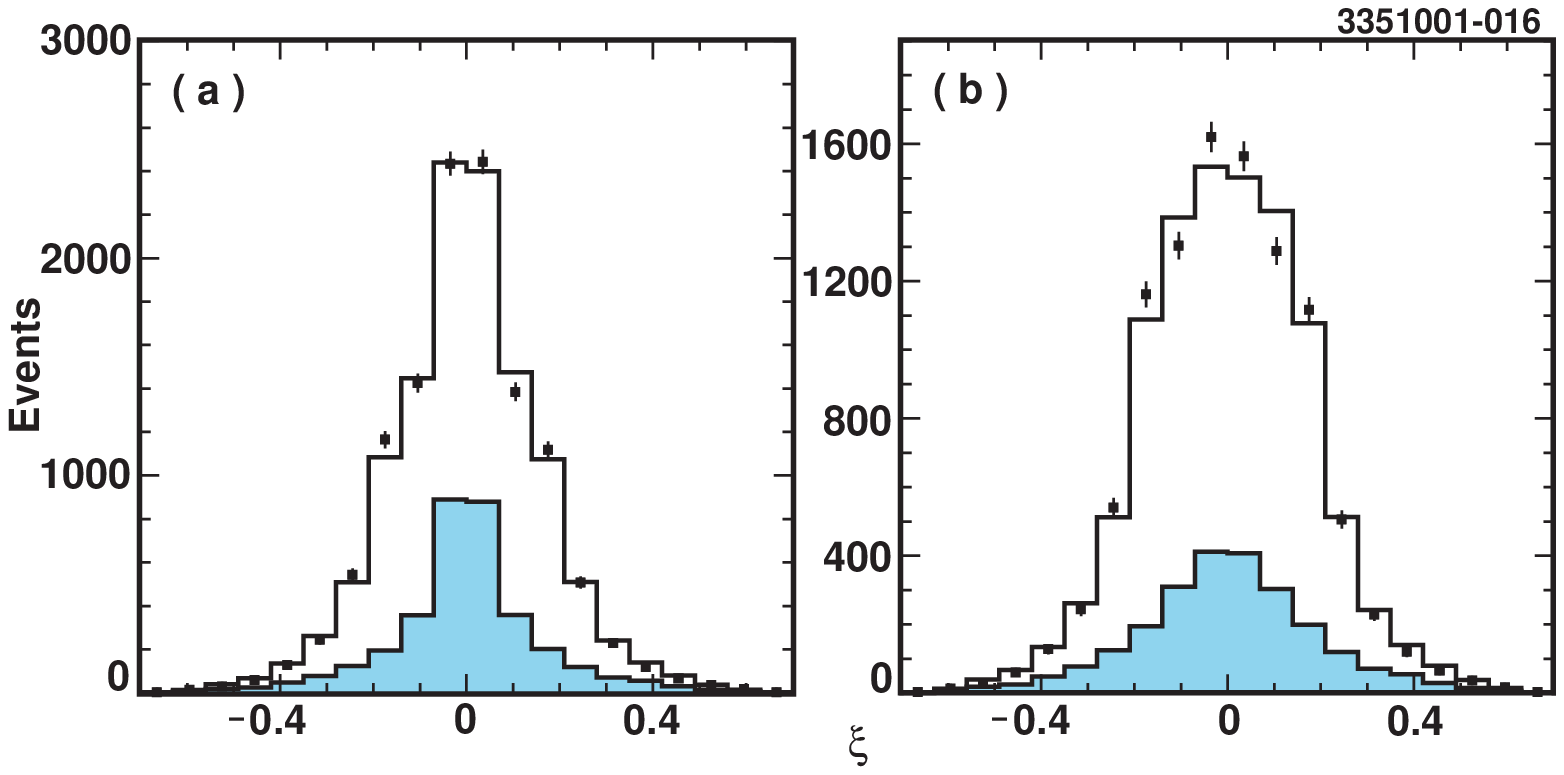}{data}{The distribution
of $\xi$ for the data (squares) compared to the sum of the background 
(hatched histogram) and a Standard Model Monte Carlo prediction for the 
(a) whole data sample and (b) for the events with the mass of the $(K\pi)$ 
system ranging between 0.85 and 1.45 GeV/$c^2$.}

\par
To relate the observed mean value of the optimal observable $<\xi>$ to
the $CP$-violating imaginary part of the coupling constant $\Lambda$, 
the $\Im(\Lambda)$ dependence of $<\xi>$ must be known.
The optimal observable is a pure $CP$-odd quantity, therefore, its average
value can be expanded in odd powers of the $CP$-odd part of $\Im(\Lambda)$. 
We have analyzed a Standard Model Monte Carlo sample with twice as
many events as the data, finding an average value of $\xi$ consistent
with zero (Table~\ref{tab:ave}). Therefore, the selection
criteria do not introduce artificial $CP$ violating asymmetry. A study of the 
$CP$-violating Monte Carlo sample showed that at small values of 
$\Im(\Lambda)$ only the few first terms in the expansion contribute to 
the $<\xi>$
\begin{equation}
<\xi> \simeq c_1 \Im(\Lambda) + c_3 \Im(\Lambda)^3.
\label{eq:xi_exp}
\end{equation} 
We estimate $c_1$ and $c_3$ from the Monte Carlo generated with different 
values of $\Im(\Lambda)$ by fitting the observed average values of $\xi$ as a 
function of $\Im(\Lambda)$ to a cubic polynomial. The obtained coefficients 
$c_1$ and $c_3$ of Eq.~(\ref{eq:xi_exp}) are given in Table~\ref{tab:c1c3}.
\begin{table}[htb]
\caption {\label{tab:c1c3}
Proportionality coefficients $c_1$ and $c_3$ for both full sample and
the region of $(K\pi)$ invariant mass ranging from 
0.85 GeV/$c^2$ to 1.45 GeV/$c^2$.}
\begin{center}
\begin{tabular}
{|c|c|c|} 
{Coefficient}&   {Full sample}  & {0.85 GeV/$c^2 < M(K\pi) < 1.45$ GeV/$c^2$}
                                                                 \\ \hline 
$c_1$        & $ 0.0368\pm0.0018$   & $ 0.0410\pm0.0020$         \\
$c_3$        & $-0.0135 \pm 0.0019$ & $-0.0127\pm0.0022$         \\ 
\end{tabular}
\end{center}
\end{table}
\par
We use these empirically determined coefficients to estimate the
value of $\Im(\Lambda)$. In Table~\ref{tab:results} we list the results 
for both the full sample and the events with restricted $(K\pi)$ invariant 
mass range. 
\begin{table}[htb]
\caption {\label{tab:results}
The values of $\Im(\Lambda)$ as well as 90\% confidence limits
for both full sample and the region of $(K\pi)$ invariant mass ranging from 
0.85 GeV/$c^2$ to 1.45 GeV/$c^2$.}
\begin{center}
\begin{tabular}
{|c|c|c|} 
{Results}&   {Full sample}  & {0.85 GeV/$c^2 < M(K\pi) < 1.45$ GeV/$c^2$}
                                                              \\ \hline 
$\Im(\Lambda)$ & $ -0.054\pm0.049$   & $-0.046\pm0.044$    \\
90\% confidence limits 
               & (-0.134, 0.027) & (-0.119, 0.027)            \\ 
\end{tabular}
\end{center}
\end{table}

\par
To estimate the upper limit on the $CP$ violating parameter $\Im(\Lambda)$ 
we first must estimate systematic errors. There are several possible sources 
of systematic
errors that can contribute to this analysis. We treat these errors to be 
multiplicative if the source can modify the value of $c_1$ and to be 
additive if the source can bias the central value of $<\xi>$.  
\par
To construct the optimal observable we parameterize the scalar hadronic
current as a product of a Breit-Wigner shape of the poorly known
scalar resonance $K^*_0(1430)$ and a normalization constant $M=1$ GeV/$c^2$
[Eq.~(\ref{eq:A})]. This assumption is a source of a systematic error on the 
calibration coefficient $c_1$. We perform the following studies on the Monte 
Carlo simulation to estimate this error. We vary the width of the $K^*_0(1430)$
resonance by $\simeq 5\sigma$ and re-calculate the coefficient $c_1$. 
This results in a change of the value of $c_1$ of $\pm4.4\%$. Similarly, we 
change the mass of the $K^*_0(1430)$ resonance from 
1.35 GeV/$c^2$ to 1.45 GeV/$c^2$ and obtain a variation of the value of $c_1$
of $\pm11\%$. Thus, the overall conservative estimate of the systematic 
error due to the uncertainty in the mass and the width of $K^*_0(1430)$ is 
$\pm12$\%.
\par
The choice of $M=1$ GeV/$c^2$ is simple but not unique. Any Lorentz invariant
quantity which has units of mass can be used as a normalization parameter.
Another choice for $M$ is the invariant mass of the $(K\pi)$ system. We
generate a Monte Carlo sample with the $(K\pi)$ mass as a normalization 
parameter and re-estimate the value of $c_1$ to be $(0.0326 \pm 0.0003)$. 
The resulting value of $c_1$ differs from its nominal value by 2\%. 
The overall 
multiplicative error on $c_1$ due to a choice of the scalar current 
parameterization is $\pm12$\%. 
\par
In Eq.~(\ref{eq:A}) we define a Standard Model $W$ exchange to have only
a pure transverse vector current. However, chiral perturbation theory
demands a scalar component~\cite{tsai_kpi}.
It is assumed to be small and is neglected in TAUOLA. To estimate the
systematic error due to the modeling of the $W$ current we modify
TAUOLA to include a scalar part. We use a Breit-Wigner shape for the $K^*_0(1430)$
to describe the scalar component. The new value differs by $\simeq~3\%$ from the 
nominal value. We take this as an estimate of the multiplicative systematic 
error due to the parameterization of the vector current.
\par
We estimate the values of the coefficients $c_1$ and $c_3$ in 
Table~\ref{tab:c1c3} from the Monte Carlo simulation. Therefore, 
the quality of the simulation may affect the result. We study the momentum 
distributions of the pion and of the reconstructed $K^0_S$ candidate in the 
signal $\tau$ decay, both in data and in the Monte Carlo. We estimate a systematic
error on the result by re-calculating the coefficients $c_1$ and $c_3$ for the 
deviations between data and Monte Carlo parameterized as a slope of the ratio 
of the momenta distributions for the real and generated data. 
The multiplicative systematic error due to imperfect simulation of the data is 
estimated to be 9.3\%.
\par
We study the systematic effects due to a possible difference of the track
reconstruction efficiency for $\pi^+$ and $\pi^-$ as a function of the pion
momentum. To estimate the size of this effect, we study the momentum 
distribution for charged pions in $\tau^\pm \to K^0_S \pi^\pm \nu_{\tau}$ 
decay. The ratio of these distributions for $\tau^+$ and $\tau^-$ decays
is consistent with 1, and the maximum deviation characterized by a slope is 
fitted to be $0.01\pm0.04$. The introduction of such a slope to the data 
sample changes the value of $\Im(\Lambda)$ by $\pm 0.009$. We take this 
as a measure of an additive systematic error.
\par
A possible source of a bias is an asymmetry of the $\xi$ 
distribution induced by the remaining background. If we denote the number of 
signal and background events by $S$ and $B$, then the contribution to the 
optimal observable due to the background is
\begin{equation}
\Delta<\xi> = <\xi>_B B/(S+B).
\label{eq:syst_bkg}
\end{equation}
Here, $<\xi>_B$ is the value of the optimal observable in the background.
We can estimate \\$<\xi>_B$ from the $\tau$ generic and multi-hadronic Monte 
Carlo simulations. For the $\tau$ background $B/S$ is $0.413\pm0.025$ and 
$<\xi>_B$ is estimated to be $(0.6\pm1.6)\times10^{-3}$. Therefore,
$\Delta<\xi>$ is equal to $(0.2 \pm 0.5)\times10^{-3}$. Such a change will 
modify the value of $\Im(\Lambda)$ by  $\pm0.014$. Similarly, the background 
contribution from the multi-hadronic processes is estimated to be $\pm0.009$. 
Therefore, the overall background contribution can modify the central value of 
$\Im(\Lambda)$ by $\pm0.017$. 
\par
The overall multiplicative error is estimated to be $\pm15$\%, and the overall
additive error on $\Im(\Lambda)$ is $\pm0.019$.
Within our experimental precision we observe no significant asymmetry of the
optimal observable and, therefore, no $CP$ violation in 
$\tau\to K\pi\nu_{\tau}$ decay. For a restricted range of the $(K\pi)$ mass
(between 0.85 and 1.45 GeV/$c^2$) we obtain a value of the imaginary part of 
the scalar component in the $\tau$ decays as
\begin{equation}
\Im(\Lambda) = (-0.046 \pm 0.044 \pm 0.019)(1 \pm 0.15).
\label{eq:final_result}
\end{equation}
The first error is statistical and the second is additive systematic. The
overall expression is multiplied by the multiplicative systematic error.
The corresponding limits are
\begin{equation}
-0.172 < \Im(\Lambda) < 0.067, \mbox{ at 90\% C.L.}.
\label{eq:final_result_CL}
\end{equation}
This limit is an order of magnitude more restrictive than that obtained
in the previous search~\cite{colin} for $CP$ violation in 
$\tau\to K\pi\nu_{\tau}$ decays. These results constrain the value of
$\Im(\Lambda)$ at a comparable level to those from our study of
$\tau^-\tau^+ \to (\pi^-\pi^0\nu_{\tau})
(\pi^-\pi^0\bar{\nu}_{\tau})$~\cite{cpv_rho}.
However, the current result is about a factor of 10 more restrictive on the 
$CP$-violating parameters of Multi-Higgs-Doublet Models~\cite{mhdm}. 
\par
We gratefully acknowledge the effort of the CESR staff in providing us with
excellent luminosity and running conditions. This work was supported by 
the National Science Foundation, the U.S. Department of Energy,
the Research Corporation, and the Texas Advanced Research Program.

\end{document}